\documentclass[10pt,sigconf,letterpaper,nonacm]{acmart}

\usepackage{hyperref}

\usepackage{caption}
\usepackage{subcaption}

\usepackage{changepage} 
\usepackage{booktabs, multirow} 
\usepackage{array,graphicx}

\begin{document}

\title[Measuring risks inherent to our digital economies]{Measuring risks inherent to our digital economies using Amazon purchase histories from US consumers}


\author{Alex Berke}
\email{aberke@mit.edu}
\orcid{0000-0001-5996-0557}
\affiliation{%
  \institution{MIT Media Lab}
  \country{USA}
}

\author{Kent Larson}
\email{kll@mit.edu}
\orcid{0000-0003-4581-7500}
\affiliation{%
  \institution{MIT Media Lab}
  \country{USA}
}

\author{Sandy Pentland}
\email{pentland@mit.edu}
\orcid{0000-0002-8053-9983}
\affiliation{%
  \institution{Stanford HAI}
  \country{USA}
}

\author{Dana Calacci}
\email{dcalacci@psu.edu}
\orcid{0000-0002-9552-1137}
\affiliation{
    \institution{Penn State University}
    \country{USA}
    }



\begin{abstract}

What do pickles and trampolines have in common? In this paper we show that while purchases for these products may seem innocuous, they risk revealing clues about customers' personal attributes — in this case, their race.  

As online retail and digital purchases become increasingly common, consumer data has become increasingly valuable, raising the risks of privacy violations and online discrimination. This work provides the first open analysis measuring these risks, using purchase histories crowdsourced from (N=4248) US Amazon.com customers and survey data on their personal attributes. With this limited sample and simple models, we demonstrate how easily consumers' personal attributes, such as health and lifestyle information, gender, age, and race, can be inferred from purchases. For example, our models achieve AUC values over 0.9 for predicting gender and over 0.8 for predicting diabetes status. To better understand the risks that highly resourced firms like Amazon, data brokers, and advertisers present to consumers, we measure how our models' predictive power scales with more data. 
Finally, we measure and highlight how different product categories contribute to inference risk in order to make our findings more interpretable and actionable for future researchers and privacy advocates.

\end{abstract}

\settopmatter{printfolios=true}

\maketitle

\section{Introduction}

Making every-day purchases increasingly requires participating in a digital economy where purchase information is collected by retailers, credit card companies, and third parties. This has created a secondary market of valuable data on consumption patterns, including purchase histories that detail the items or categories consumers purchase over time. This data is part of a multibillion dollar data market helping firms infer personal attributes about consumers, highly valuable for ad targeting and predictive analytics~\cite{knowledgesourcingintelligencellp}. 

The consumer attributes revealed from purchases can also be sensitive, posing significant privacy risks, as highlighted by high-profile journalistic reports~\cite{duhigg2012}. 
Yet while these risks are not new, their past descriptions are qualitative or anecdotal and researchers have lacked open access to data needed for quantitative measurements.

In this paper we present the first open quantitative analysis measuring the risks purchases data pose to US consumers by leveraging a recently published dataset of purchases from Amazon, the largest online retailer in the US~\cite{statista2023}.
The dataset contains purchase histories from thousands of consumers, along with their self-reported demographic, health and lifestyle information.
We use this data to demonstrate the ease of inferring sensitive consumer traits from purchases, including gender, age, race, and whether consumers have diabetes or use alcohol, cigarettes or marijuana.

Although this novel dataset allows us to demonstrate inference risks, it is small compared to data available to Amazon and other companies. 
For this reason, our models are intentionally simple, measuring the lower bounds of inference risks, and we show how the models' predictive performance increases with more data.
To provide greater intuition into how these data can be so revealing, we then measure which purchase categories contribute most to inference risk across attributes. 
These contributions demonstrate the privacy risks due to purchases data available to companies via our increasingly online economies, how risk scales with dataset size, and how different types of purchases present most risk to consumers.

\section{Background and related work}

\subsection{Amazon's growing market dominance}
\label{section:background:amazon}

By 2018 more than 60\% of US consumers had bought an item through Amazon~\cite{npr/maristpoll2018} and by 2024 more than 50\% (180 million) were Amazon Prime (subscription service) members~\cite{bloomberg2024}. Here we highlight substantial investments Amazon made to acquire its significant market share, and hence shopper data, which we revisit in analyses.

Amazon began by selling books, strategically acquired other book sellers, and by 1998 referred to itself as "Earth's Biggest Bookstore"~\cite{secinfo98}.
In 2009 Amazon acquired the leading online footwear company, Zappos~\cite{zapposSEC2009}, and in 2017 Amazon acquired Whole Foods Market Inc. for \$13.7 billion, where reporters attributed this large spend as an opportunity to acquire data~\cite{wsjWF2017}. 
Beyond acquisitions, Amazon has invested in creating a platform where third-party sellers offer a broader array of products than any one company could manufacture or source~\cite{weigel2023}, providing an even broader array of purchases data. This was highlighted in 2020 by Amazon's CEO:  "Third-party sales now account for approximately 60\% of physical product sales on Amazon, [...] growing faster than Amazon’s own retail sales"~\cite{amazonstaff2020}.

Amazon has also made acquisitions in the healthcare space, acquiring the online pharmacy PillPack in 2019~\cite{cnbcPillPack2019} and One Medical, a chain of primary care clinics, in 2022.  
Again, reporters described this as a data opportunity~\cite{nytimesOneMed2022}.  Although HIPAA rules~\cite{hipaa} protect health records from the rest of the Amazon business, we ask if there are risks due to the lack of protection in the other direction. Could purchases data be used to help determine how medical services are targeted? In this work we show how Amazon purchases can reveal health information, such as whether someone has diabetes.

\subsection{Demographics and inference}

Related work shows the utility of demographic data in predicting purchases and informing business models~\cite{islam2022}, including studies specific to online retail~\cite{hood2020, dominici2021, hamad2019}. Given the utility of consumer demographics, inferring demographics from purchases is also an active area of research, where the goal is to provide businesses with more data on their customers, to improve their market strategy~\cite{wang2016, kim2019, jiang2019}. These previous studies have used data provided by retailers~\cite{wang2016} and vendor loyalty program providers~\cite{kim2019, jiang2019}.
Our analyses also infer consumer demographics from purchases data. 
Yet in contrast, our goal is to measure related privacy risks using data crowdsourced from consumers, rather than optimizing model utility or accuracy.

There is also related work analyzing the privacy risks of purchases data, where the focus is on re-identification risks~\cite{deMontjoye2015, li2023}.
For example, De Montjoye et al. used 3 months of credit card records to show how easily users could be re-identified in a de-identified dataset, and showed risks were higher for women versus men~\cite{deMontjoye2015}. Further research suggests re-identification risks are understated in marketing datasets due to their longitudinal nature~\cite{li2023}.

Other work shows how different types of users' digital traces, such as social media Likes~\cite{kosinski2013} or web browsing histories~\cite{ourFlocPaper}, can be used to infer sensitive sociodemographics. In an influential 2013 paper, Kosinski et al. used a dataset of 58k Facebook users' Likes and demographic profiles to show how well a model could infer users' personal attributes from the Likes, reporting AUC scores of 0.93 for predicting gender, and 0.73 for cigarette use~\cite{kosinski2013}. Our paper adds to this body of work by highlighting similar risks with purchases.

\section{Data and modeling approach}

\begin{table*}
\caption{Example purchases data: A representative sample of rows from one user's Amazon data.}
\begin{adjustwidth}{-7.75cm}{-7.75cm} 
\label{tab:purchases_data_example_rows}
\centering
\small
\begin{tabular}
{llp{6.7cm}lll}
\toprule
\textbf{Date} & \textbf{Price} & \textbf{Title} & \textbf{Product code} & \textbf{Category} & \textbf{ResponseID} \\
\midrule
2018-01-21 & \$23.07 & OTTERBOX SYMMETRY SERIES Case for iPhone 8& B01K6PBRSW & CELLULAR\_PHONE\_CASE & R\_2zARigFdY \\
2018-02-06 & \$15.91 & The Power of Transcendental Meditation & 1501161210 & ABIS\_BOOK & R\_2zARigFdY \\
2018-04-03 & \$5.99 & Square Reader for magstripe (with headset jack) & B00HZYK3CO & MEMORY\_CARD\_READER & R\_2zARigFdY \\
2018-06-11 & \$4.89 & Dove Advanced Care Antiperspirant Deodorant Stick for Women, Original Clean, for 48 Hour Protection & B00Q70R41U & BODY\_DEODORANT & R\_2zARigFdY \\
\bottomrule
\end{tabular}
\end{adjustwidth}
\end{table*}

Our analyses use a dataset of purchase histories collected from US Amazon users, along with their demographic, health and lifestyle information reported through a survey.
The data were collected and published with users' informed consent, as described in~\cite{ourCSCWpaper}. The dataset was previously detailed and published via~\cite{ourSciDataPaper}. 
Note that while this is the first such dataset available for open research, companies have access to even larger datasets linking purchases and demographics. Consider Nielsen's consumer panel tracking over 250k households~\cite{nielsenconsumerllc2022} and Amazon's own shopper panel~\cite{amazonShopperPanel}.

\subsection{Purchases data and preprocessing}

Table \ref{tab:purchases_data_example_rows} shows a representative sample of one user's Amazon data. There is a row for each purchase with an order date, unit price, product title, product code (ASIN/ISBN), and category. A "ResponseID" added by the data collection software links purchases to users.

\subsubsection{Categories and preprocessing}

\label{section:categories preprocessing}

The following analyses use the product categories, assigned by Amazon, rather than product codes/titles because the data are sparse in products, with many products purchased by only a few users.
The product categories provide a natural clustering over products, which we use instead of algorithmic clustering, given our goal to provide both quantitative and interpretable results.
To help overcome limitations of using categories rather than the rich information available to Amazon in product codes/titles, we make the following modifications.

Gendered products:
Given gender is often explicit in product titles but not categories, using categories without these gendered labels risks misrepresenting the information leaked from purchases. To fix this, we prefix each product's category with "MEN"/"WOMEN" if Men/Women is in the product title.

Books:
The categories for books in this dataset lack the rich genre information that is publicly available via their titles and product codes.
E.g. before preprocessing, "ABIS\_BOOK" is the category for more than 95\%. We update book categories by pulling categories from the Google books API. For books where the API returned a category, C, we assign the category "BOOK:C". See the Appendix~\ref{appendix:data and preprocessing} for details.

\subsubsection{Data restrictions}
\label{section:data restrictions}

We restrict the data to users who purchased at least 50 different products, and categories purchased by at least 50 different users. After preprocessing, our sample includes 1,802,908 purchases, 910,474 unique products, 1619 categories, and N=4248 users. 
See Appendix~\ref{appendix:metrics distributions} for data distributions.

\subsection{Demographics, health and lifestyle data}

We use the following self-reported user attributes: gender, age group, race, and answers to questions about health and lifestyle. 
To conduct analyses with sufficient sample size, we limit gender analyses to the Male/Female binary, excluding the small number of "Other" users, and limit race analyses to the categories White, Black, and Asian. Some users reported multiple races. We positively label users for each race group they reported, allowing users to be in multiple race groups.

Survey participants also answered the following questions about themselves or anyone in their household or that they shared their Amazon account with: 
"Smoke cigarettes regularly?", "Smoke marijuana regularly?", "Drink alcohol regularly?", for which they could answer (Yes/No/Prefer not to say/Recently stopped), and "Have diabetes?" (Yes/No/Prefer not to say). Our analyses only use the Yes/No answers to these questions.

Table \ref{tab:sample_demos_n} shows the number of users with each demographic and health/lifestyle attribute. Note that due to how some values are excluded from analysis (e.g. gender: "Other") or counted (i.e. race), the number of users in each group (n) does not necessarily sum to the total sample size (N).

\begin{table}[]
\caption{Sample descriptive statistics.}
\label{tab:sample_demos_n}
\centering
\small
\begin{tabular}{p{1cm}llll}
\toprule
\multicolumn{2}{l}{\textbf{Attribute}} &\multicolumn{2}{l}{\textbf{n (N=4248)}} \\\midrule
\multirow{2}{*}{\textbf{Gender}} &Male &1847 & \\
&Female &2301 & \\
\hline
\multirow{3}{*}{\textbf{Age}} &18 - 34 years &2084 & \\
&35 - 54 years &1694 & \\
&55 and older &470 & \\
\hline
\multirow{3}{*}{\textbf{Race}} &White &3555 & \\
&Asian &376 & \\
&Black &354 & \\
\midrule
\multicolumn{5}{p{7.5cm}}{\emph{"Are any of the following the case for you or someone in your household or someone you share your Amazon account with?"}} \\
\hline
& &\textbf{Yes} &\textbf{No} \\
\hline
\textbf{Health}&Diabetes &516 &3720 \\
\textbf{\&}&Cigarettes &600 &3506 \\
\textbf{lifestyle}&Marijuana &892 &3215 \\
&Alcohol &1874 &2229 \\
\bottomrule
\end{tabular}
\end{table}

\subsection{Modeling approach}

We use a simple modeling approach to demonstrate the ease of inferring sensitive consumer attributes from purchases data. Our goal is to demonstrate lower bounds of this risk, rather than to develop optimal models.

\subsubsection{Feature vectors}

We transform purchase histories into simple feature vectors by one-hot-encoding product categories. For a given category, 1 indicates the user bought a product in that category without indicating how often. 

\subsubsection{Models and training versus testing data}
We test the risk of inferring each attribute listed in Table~\ref{tab:sample_demos_n} separately: we treat each attribute as a label in a separate binary prediction task.  For each attribute, a user is assigned a label of 1 if they have the attribute, 0 otherwise.  We make a separate 80/20 train/test split for each attribute, stratified on the label, which is consistently used for all modeling tasks for the given attribute.
We train a Support Vector Machine with the radial basis function (RBF) kernel and balanced class weights~\cite{sklearn2011} for each attribute. We evaluate models by their area under the receiver operator curve (AUC)~\cite{bradley1997}. We use AUC because it provides an aggregate measure of performance across all possible classification thresholds.

\section{Inference results}
Given our limited dataset and simple models, all results should be considered lower bounds.

Table~\ref{tab:auc_results} shows the model results (AUC) for predicting each attribute. AUC scores higher than 0.5 represent predictive power, 1 represents perfect prediction. The models demonstrate predictive power for all attributes and show how some are easier to infer than others. For example, gender is relatively easier to predict, older consumers are most at risk of age inference, and whether consumers are Black is easier to infer versus White or Asian. This is the case even though older and Black consumers are not as well represented in the data, providing fewer training samples.

The models also show predictive power for health and lifestyle attributes: models predicting diabetes status or regular use of cigarettes, marijuana, or alcohol yielded AUC scores well above 0.5. A limitation of these results is that the questions about health and lifestyle were asked about an entire household. If another household member regularly used cigarettes/marijuana/alcohol, but the survey participant did not, they were still positively labeled. Similarly for diabetes.  This resulting disconnect between label and personal purchases likely contributed noise to the data, resulting in lower AUC.
To account for this, we leverage survey data on participants' household sizes and whether they share their Amazon count. We repeat the analyses after restricting data to users who are the sole members of their household and do not share their accounts. Resulting AUC for predicting diabetes climbs to 0.819. This reduction of statistical noise also impacts inference for other demographics. AUC scores for male and female increase to 0.962 and 0.968, respectively. Details on this additional analysis are in the Appendix (\ref{appendix:single purchasers}).

\begin{table}[]
\centering
\caption{Model results.}
\label{tab:auc_results}
\small
\begin{tabular}{p{1cm}lrr}\toprule
\multicolumn{2}{l}{\textbf{Attribute}} &\textbf{AUC} \\\midrule
\multirow{2}{*}{\textbf{Gender}} &Male &0.893 \\
&Female &0.912 \\
\hline
\multirow{3}{*}{\textbf{Age}} &18 - 34 years &0.808 \\
&35 - 54 years &0.712 \\
&55 and older &0.868 \\
\hline
\multirow{3}{*}{\textbf{Race}} &White &0.791 \\
&Black &0.806 \\
&Asian &0.764 \\
\hline
\multirow{3}{2cm}{\textbf{Health\\\&\\ lifestyle}} &Diabetes &0.612\\
&Cigarettes &0.728 \\
&Marijuana &0.634 \\
&Alcohol &0.631 \\
\bottomrule
\end{tabular}
\end{table}

\subsection{Increasing data and model performance}

Consider how Amazon continues to increase the categories of consumers' purchases available to the company via acquisitions and third-party sellers. This increase in categories generalizes to other data markets, where data brokers and advertisers acquire increasing sources of data.

This section measures the impact of increasing both the breadth and quantity of purchase categories on predictive performance. 
We do this by creating data samples limited to a subset of categories and then show how expanding the data to include more categories improves model results.
For each subset of categories, we train and test models limited to purchases in those categories, using the same train/test split as the other analyses.
Note that limiting the categories has the effect of limiting both category diversity and total purchases, where many users may have few or no purchases for a given subset of categories.

When subsampling categories, we first demonstrate a random sampling approach. We then present analysis where category groups are added sequentially to parallel the high profile acquisitions by Amazon described in Section~\ref{section:background:amazon}.

\subsubsection{Randomly sampled categories analysis}

\begin{figure}
    \centering
    \includegraphics[width=0.86\linewidth]{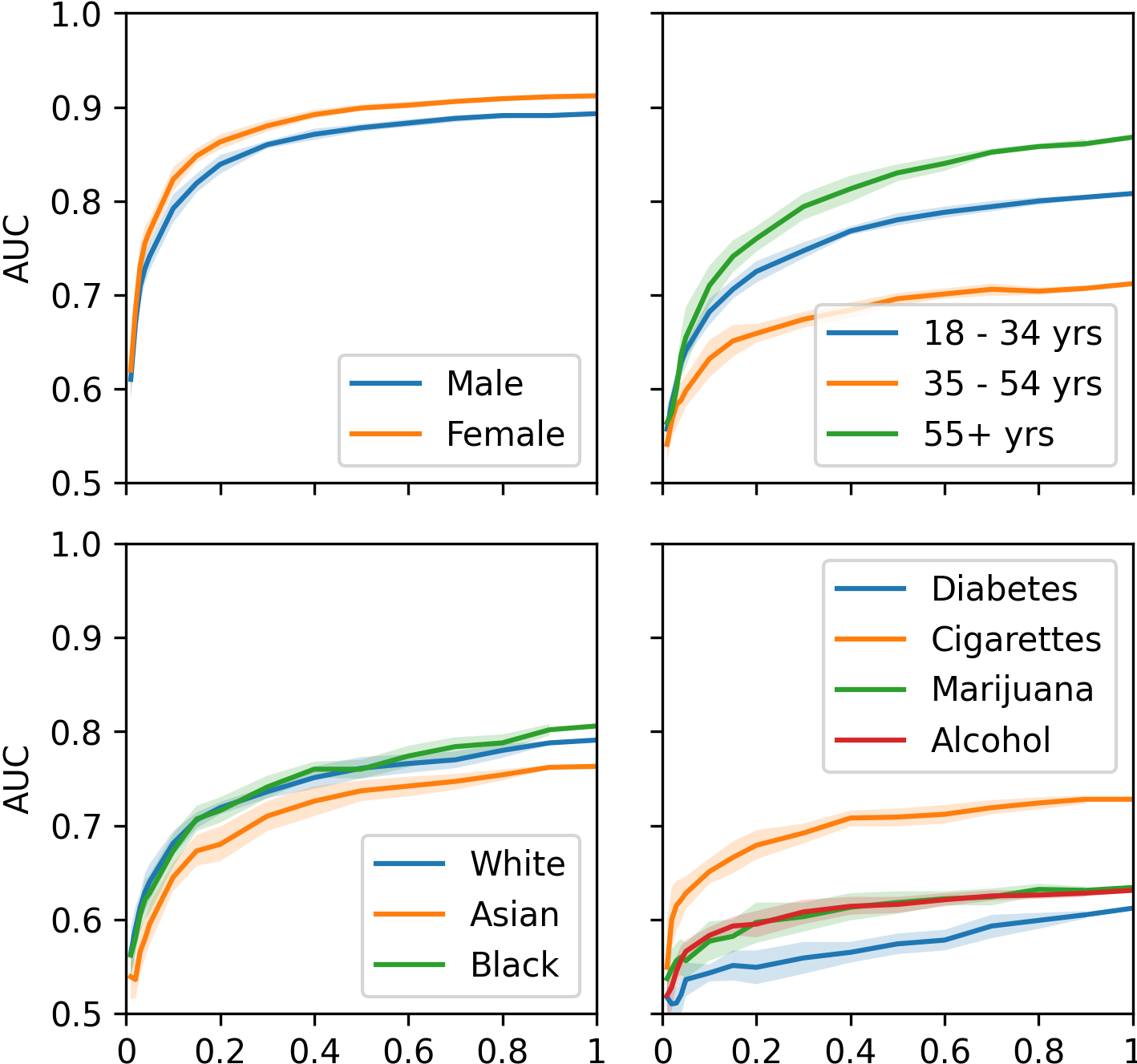}
    \caption{Prediction improvements with increasing fraction of categories (x-axis) available to model.}
    \label{fig:auc_results_by_cat_additions_rand_fraction}
\end{figure}

For $p$ ranging from 0.01 to 1, we randomly sample a fraction of $p$ categories, without replacement. We then recreate the feature vectors using this subset of categories, and retrain and test the models. We repeat this random sampling process to compute means and confidence intervals for the resulting AUC values. 
The mean AUC values and 95\% CIs are shown in Figure~\ref{fig:auc_results_by_cat_additions_rand_fraction} with
numeric results in Appendix~\ref{appendix:changes with data}.

Results show how some attributes, such as gender, can be easily predicted with a small portion of data, where more data yields diminishing returns.
For example, AUC for male and female are above 0.6 when using only 1\% of the categories, and above 0.8 when using only 15\%.  For other attributes, results suggest that more data, beyond our dataset's limitations, would further improve predictive power: AUC continues to slope upward even as the categories available in our dataset are exhausted. This is particularly the case for the 55 and older group, race groups, and diabetes status.

\subsubsection{Acquisitions and thematically sampled categories}

\begin{figure*}
    \centering
    \includegraphics[width=0.86\linewidth]{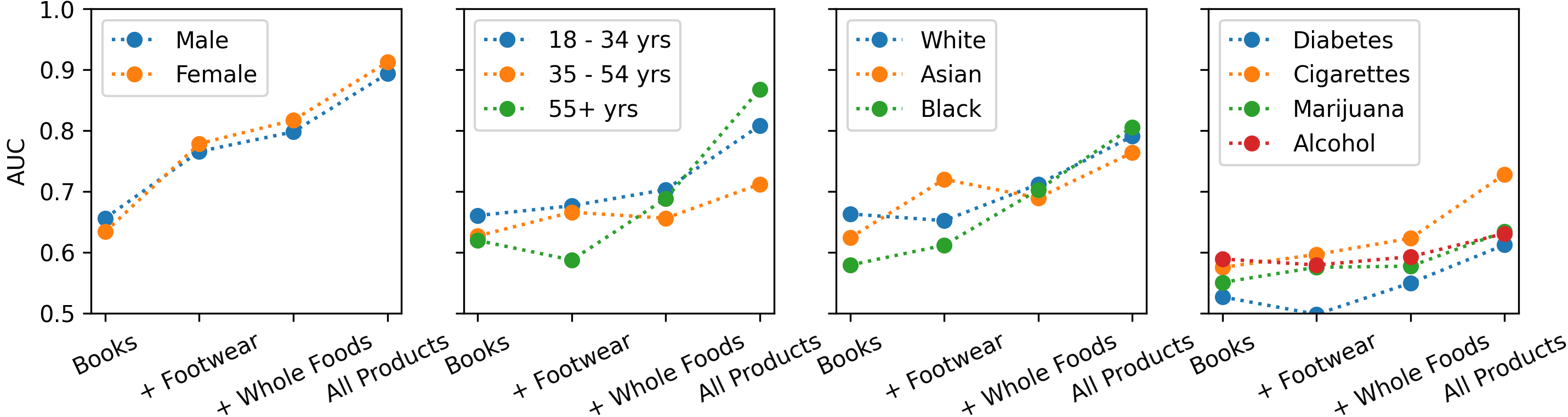}
    \caption{Change in model performance with additional product categories related to Amazon acquisitions.}
    \label{fig:auc_results_by_cat_acquisitions}
\end{figure*}

We first limit the data to books and then add groups of product categories in a sequence parallel to Amazon's high profile acquisitions (see Section~\ref{section:background:amazon}): we add footwear, then Whole Foods products, then all other products available to Amazon.

Books:
We identify books and related products by identifying product codes that are ISBNs~\cite{internationalisbnagency2024}. 
This includes 190 categories and 87,157 purchases.

Footwear:
We add footwear purchases by filtering to categories matching any of  `SHOES', `TECHNICAL\_SPORT\_SHOE', `BOOT', `SANDAL', `SLIPPER', `SOCKS', including gendered versions of these categories, as labeled in data preprocessing. This results in 18 additional categories.

Whole Foods:
Although Whole Foods offers a vast catalog of groceries, we simply add products that have "Whole Foods" in the title, resulting in 119 additional categories. Some of these categories are gendered, namely vitamins.

Results are shown in Figure~\ref{fig:auc_results_by_cat_acquisitions} with numeric results in Appendix Table~\ref{tab:auc_results_by_cat_acquisitions}.
Results show that books alone provide predictive power for demographic groups and how the further additions of product categories have different impacts on different demographics. For example, adding footwear purchases, where many products are explicitly gendered, improves predictive power for gender, but not necessarily for other demographics, and instead might add statistical noise. Adding product categories beyond books, footwear and Whole Foods products was important in improving the predictive power of models, particularly for predicting the oldest age group and health and lifestyle attributes.

\section{Categories contributing to risk}
\label{section:categories contributing risk}

This analysis measures how different purchase categories impact inference risks, highlighting categories that are both positive and negative predictors, with the goal of providing interpretable results.

\subsection{Methodology}

We train a logistic regression model to predict each demographic and health/lifestyle attribute, using the same feature vectors as the previous analyses.
We use the estimated coefficients to compute the odds ratio (OR) associated with each category. OR greater than 1 indicates a category is a positive predictor for a given attribute, where higher values are more predictive; a value less than 1 indicates a category is a negative predictor, where smaller values have more negative impact.
We use logistic regression for this analysis, versus SVM, because the resulting ORs provide a more interpretable meaning of how much each category impacts inference and can reflect both positive and negative predictors. AUC results for the logistic regression and SVM are similar, slightly higher for the SVM (see Appendix Table~\ref{tab:lr_auc_results}).

\subsection{Results}

\begin{figure}
\centerline{\includegraphics[width=0.68\linewidth]{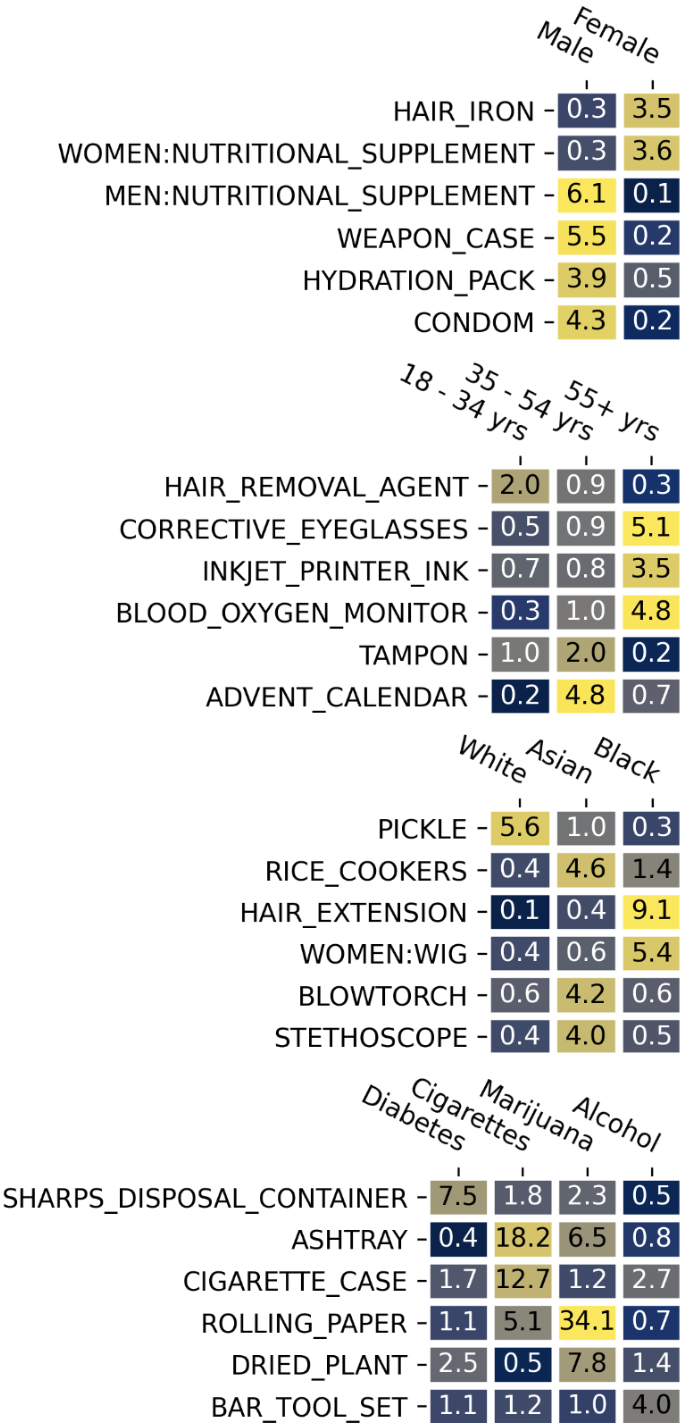}}
    \caption{Categories with estimated odds ratios.}
    \label{fig:categories_risk}
\end{figure}

Figure~\ref{fig:categories_risk} highlights a sample of results with heatmaps coloring categories by their ORs. 
These categories were chosen from the top 10 most positive and negative predictors for each attribute. See Appendix~\ref{appendix:categories contribution to risk} for more categories and ORs.

In many cases, a positive predictor category for one demographic group is a negative predictor for another. We highlight an obvious example: women's and men's nutritional supplements are positive predictors for females and males, respectively, and negative predictors for the opposite gender. Other categories are less obvious, e.g. pickles are positive predictors for White and negative predictors for Black consumers.
While many of these results may seem intuitive, consumers may not realize, or have the agency to change, how their purchases impact inference risks.
For example, corrective eyeglasses are positive predictors for the 55 and older group, and tampons are negative predictors.
For the health/lifestyle attributes, there are predictive categories with substantially higher OR values. For example, sharps disposal containers, commonly used by diabetics, have an OR greater than 7 for predicting diabetes. Ashtrays and cigarette cases have ORs greater than 12 for predicting cigarette use, and rolling paper has an OR greater than 30 for marijuana use.
This is in contrast to the demographic attributes, where there are less dominant predictors and instead many predictive categories with individually less predictive power.

\section{Discussion}

Our results show how consumers inadvertently risk revealing sensitive information to retailers, advertisers, and data brokers by simply making online purchases. 
With a small sample and simple models we measured the ease with which consumers' demographics can be inferred, with the lowest AUC values above 0.7 and the highest above 0.9. Given our data limitations compared to data available to Amazon and other companies, we present our results as lower bounds.

\subsection{Risks to consumers}
\subsubsection{Health and lifestyle information}

Our results show how sensitive health and lifestyle information, such as alcohol, cigarette, marijuana use and diabetes status, can be inferred from purchases. 
This was the case even when using noisy data, where consumers reported these attributes about anyone in their households.
When we repeated the analysis with less noisy data, limited to consumers living alone, the model predicted diabetes status with AUC above 0.8.  
These results suggest other sensitive health information, not collected by the survey, may also be inferred from purchases. Previous reporting has highlighted the high value of information about consumers' health conditions and prescription medications for marketing~\cite{ft2013}.  Given Amazon's acquisition of PillPack, an online pharmacy, and One Medical, which provides medical services, this information could be valuable for Amazon when targeting their marketing or services. While we have no evidence that Amazon uses data in this way, we ask: would such a use of consumers' data be problematic? 
As consumers increasingly make health-related purchases online, the health and privacy implications may warrant further scrutiny from regulators and advocates.

\subsubsection{Intersectionality}

Our analyses predict consumers' attributes in isolation due to sample limitations, yet Section~\ref{section:categories contributing risk} highlights risks regarding intersectionality. For example, some women's products, e.g. wigs, are highly predictive of whether the consumer is Black, where Black woman is then implied, suggesting Black women purchasing these products may be at particularly high risk of demographic inference. Future analyses should further explore intersectionality.

\subsubsection{Increasing risks}

Our analyses also demonstrate how inference risks increase when companies have access to a broader array of purchase categories. Amazon has acquired companies and created a platform for third-party sellers and we can expect the company to continue to diversify its data acquired from consumers.   
For example, Amazon offers credit cards to Prime customers~\cite{primeVisa2017}, incentivizing sign-ups with a \$200 gift card~\cite{primeVisa2002023}. Credit card transactions can provide insights into purchases beyond those made on Amazon's marketplace, and beyond what we were able to analyze. These increasing risks should also be considered in the broader context of data economies, where data brokers and advertisers amass consumer data from broader arrays of sources.

\subsection{Implications for online ad delivery}

The implications of our results present risks for both consumers and advertisers.
The ease with which our simple models inferred sensitive attributes suggests these attributes may be latent variables in ad delivery models. This can increase risks of inadvertent bias in ad delivery systems, potentially exposing advertisers to legal liability.  

For example, ad platforms such as Facebook use machine learning models to optimize ad delivery, leveraging a variety of signals collected from users~\cite{FBmlAdDelivery}. 
A 2023 study by Consumer Reports found that on average, more than 2,200 different companies had shared data with Facebook for each study participant, with Amazon among the top 10 companies exhibiting such a data transfer~\cite{CR2023}.  The US has regulations that prohibit preferencing ads for housing~\cite{fairhousingact24CFR} or employment~\cite{titleVII1964employment} based on demographics and in 2018 Facebook was sued for alleged violations~\cite{FBfairhousingactlawsuite}. Targeting such ads based on demographics is now prohibited by the major ad platforms Facebook~\cite{FBAdPolicies} and Google~\cite{GoogleAdPolicies}. Yet researchers have demonstrated how employment ads on Facebook can still target US audiences based on protected demographics due to signals not made clear through the ad platform~\cite{kaplan2022}. 
This presents an example of how problematic demographic biases can emerge in ad delivery systems, breaking the system's own policies, potentially exposing them to legal actions. 
Given that data from Amazon and numerous other companies are transferred to ad networks like Facebook, and our findings that demographics may be latent variables in these data, our results suggest use of consumer data in ad networks may exacerbate risks of discrimination for both advertisers and consumers.

\subsection{Future work to address limitations}

This work is only a first step towards openly measuring risks in our digital economies.
Our open analysis was made possible by the open publication of a dataset crowdsourced from Amazon users~\cite{ourSciDataPaper}, yet the analysis is highly limited by the dataset's relatively small size. In order to better measure inference risks, future work can pursue crowdsourcing more data and extend our analyses with more powerful modeling.

\bibliographystyle{ACM-Reference-Format}
\bibliography{references}

\appendix

\section{Data and code availability}

All data and analysis code used in this work are available via an open repository: 
\url{https://github.com/aberke/amazon-study}.

\section{Ethics}

The Amazon purchase histories and demographics data used in this work were previously published via an IRB approved study (MIT protocol \#2205000649) with users' informed consent~\cite{ourCSCWpaper}. In our analyses we do not reveal any personal data that was not already self-reported by study participants.

\section{Data and preprocessing}
\label{appendix:data and preprocessing}

\subsection{Books}

We identified books products as those that had product codes matching the ISBN format\footnote{\url{https://en.wikipedia.org/wiki/ISBN}}: product codes with 9 to 13 all numeric characters, allowing a final `X' character.

We identified ISBNs (product codes) where leading zeros had likely been dropped in the data ingestion process and updated them in the data we used. 
This may have happened if the product codes were transformed to numbers in the data exportation process - the Amazon data were crowdsourced from study participants who exported their data from Amazon.
We did this by identifying likely ISBNs, with fewer than 10 characters, that matched ISBNs in the data after prefixing them with one or two zeros.
We updated the ISBNs where the leading zero(s) were likely dropped to their matched ISBN that started with zero(s). This resulted in 1,159 (of 59,440) updated ISBNs.

To access the categories from the Google Books API we used the following API endpoint, Where "ISBN" was the ISBN we were querying data for:

https://www.googleapis.com/books/v1/volumes?q=isbn:ISBN

We used the returned "categories" data (single item list), if available.
We then updated books categories as follows: 
We renamed the "ABIS\_BOOK" category to "BOOK".
If a Google Books API category, C, was available, we assigned the category "BOOK:C".

Table \ref{tab:top10_books_cats} shows the top 10 books categories, after this preprocessing, with frequency counts. Most common is "BOOK", the category indicating no category information was found via the Google Books API.
Table \ref{tab:books_cats_counts_stats} shows descriptive statistics for the number of occurrences, showing that most categories occurred only once.

\begin{table}[h]
    \caption{ Top 10 most common books categories.}
    \label{tab:top10_books_cats}
    \centering
    \footnotesize
\begin{tabular}{lrr}\toprule
\textbf{Category} &\textbf{Count} \\\midrule
BOOK &14167 \\
BOOK:FICTION &5687 \\
BOOK:JUVENILE FICTION &5595 \\
BOOK:JUVENILE NONFICTION &2576 \\
BOOK:COMICS \& GRAPHIC NOVELS &2281 \\
BOOK:BIOGRAPHY \& AUTOBIOGRAPHY &1609 \\
BOOK:COOKING &1303 \\
BOOK:RELIGION &1288 \\
BOOK:HISTORY &1271 \\
BOOK:BUSINESS \& ECONOMICS &1239 \\
\bottomrule
\end{tabular}
\end{table}

\begin{table}[h]
    \centering
    \caption{Descriptive statistics for value counts for the books categories.}
    \label{tab:books_cats_counts_stats}
    \footnotesize
\begin{tabular}{lrr}\toprule
count &2832 \\
mean &20.9 \\
std &320.8 \\
min &1 \\
25\% &1 \\
50\% &1 \\
75\% &2 \\
max &14167 \\
\bottomrule
\end{tabular}
\end{table}

\subsection{Categories}
\label{appendix:top categories}

Some product codes were associated with multiple categories in the dataset. We used the most frequently occurring category for each product. 
Table \ref{tab:top10_cats} shows the top 10 categories, sorted by the number of distinct users who purchased products from the category, along with the total number of purchases made from the category and average (median) cost per item. Many purchases lack an assigned category. This is indicated as "UNCATEGORIZED*" in the Table and results as the top category, as nearly all users (4203 of N=4248) purchased an uncategorized item.

\begin{table}[]
\centering
\caption{Top 10 categories by users making purchases\\(*UNCATEGORIZED includes all purchases missing a category).}
\label{tab:top10_cats}
\footnotesize
\begin{tabular}{lrrrrr}\toprule
\textbf{Category} &\textbf{Users} &\textbf{Purchases} &\textbf{Avg cost} \\\midrule
UNCATEGORIZED* &4203 &87297 &\$15.98 \\
ELECTRONIC\_CABLE &3296 &17917 &\$9.99 \\
CELLULAR\_PHONE\_CASE &3211 &14127 &\$12.99 \\
BOOK &3160 &20250 &\$11.71 \\
HEALTH\_PERSONAL\_CARE &3132 &16049 &\$11.99 \\
HEADPHONES &3100 &10958 &\$24.99 \\
CHARGING\_ADAPTER &2732 &8017 &\$14.99 \\
NUTRITIONAL\_SUPPLEMENT &2606 &24878 &\$19.01 \\
SKIN\_MOISTURIZER &2576 &13015 &\$12.95 \\
BATTERY &2532 &10331 &\$10.99 \\
\bottomrule
\end{tabular}
\end{table}

\subsection{Distributions of purchases, products, categories per user}
\label{appendix:metrics distributions}

\begin{figure}
    \centering
    \includegraphics[width=0.6\linewidth]{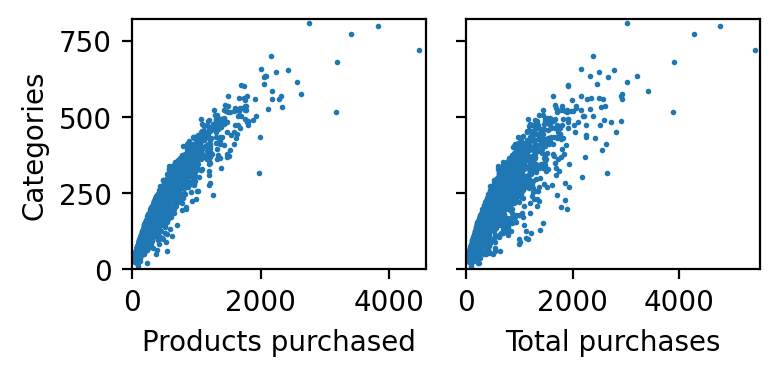}
    \caption{Number of categories purchased per user compared to the number of products purchased per user (left) and compared to the number of total purchases per user (right).}
    \label{fig:cats_vs_products_and_purchases}
\end{figure}

Table \ref{tab:per_user_purchases_stats} shows the number of purchases, products, categories per user, limited to our sample of users with at least 50 products purchased (N=4248).
Across users in the sample, the median number of total purchases is 291, the median number of categories is 142, and the median number of products is 255.5. 
Figure~\ref{fig:cats_vs_products_and_purchases} (left) shows the number of categories per user compared to the number of products purchased per user (Pearson r=0.936; p<0.001), and (right) shows the number of categories per user compared to the total purchases per user (Pearson r=0.895; p<0.001).

\begin{figure}
    \centering
    \includegraphics[width=1.03\linewidth]{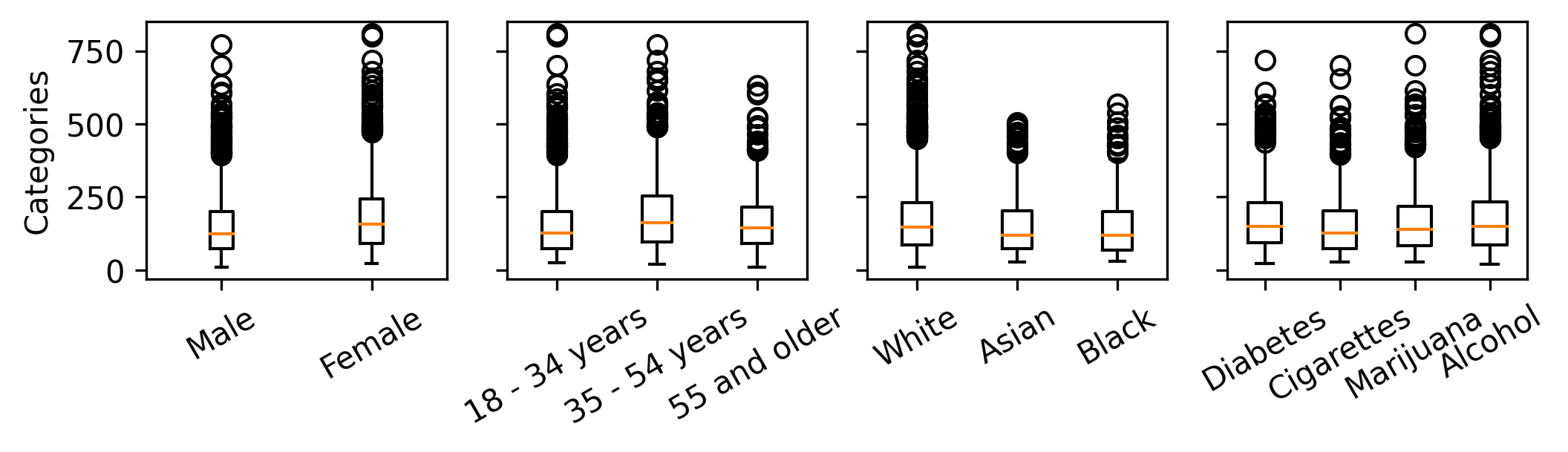}
    \caption{Number of distinct product categories purchased from, per user, for each demographic and health/lifestyle group.}
    \label{fig:cats_boxplots_by_label}
\end{figure}

The distributions of the number of purchases, products, and categories are similar across demographic groups. This is important in our analyses because our models leverage the content and diversity of purchases rather than quantity.
Figure \ref{fig:cats_boxplots_by_label} shows the distributions of the number of distinct product categories purchased from, per user, for each demographic and health/lifestyle group.

\begin{table}[]
    \centering
    \caption{Purchases, products, categories per user.}
    \label{tab:per_user_purchases_stats}
    \footnotesize
    \begin{tabular}{llll}
    \toprule
& purchases & products & categories \\
\midrule
mean & 430.2 & 360.5 & 167.9 \\
std & 436.1 & 347 & 110.8 \\
min & 50 & 50 & 10 \\
25\% & 148 & 132 & 82 \\
50\% & 291 & 255.5 & 142 \\
75\% & 551.2 & 461.2 & 225 \\
max & 5415 & 4461 & 810 \\
\bottomrule
\end{tabular}
\end{table}

\section{Inference results}

\subsection{Single versus non-single purchasers}
\label{appendix:single purchasers}

\begin{figure}
    \centering
    \includegraphics[width=0.85\linewidth]{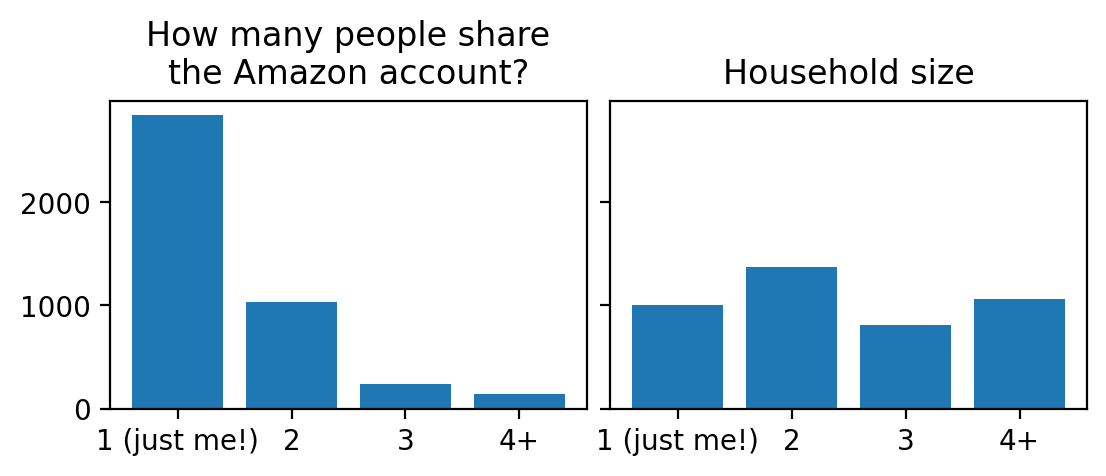}
    \caption{Response counts for questions about how many people survey participants share their Amazon account with (left) and the number of people in participants' households (right).}
    \label{fig:amzn_account_sharing_counts}
\end{figure}

\begin{table}
    \centering
    \caption{Model results for "single purchasers" versus "non-single purchasers", shown by area under the receiver operating characteristic curve (AUC).}
    \label{tab:single_purchaser_auc_results}
    \footnotesize
\begin{tabular}{lrrr}\toprule
&\multicolumn{2}{c}{AUC} \\\cmidrule{2-3}
Attribute &Singer purchaser &Non-single purchaser \\\midrule
Male &0.962 &0.869 \\
Female &0.968 &0.894 \\
18 - 34 years &0.813 &0.808 \\
35 - 54 years &0.661 &0.723 \\
55 and older &0.884 &0.855 \\
White &0.763 &0.798 \\
Black &0.785 &0.81 \\
Asian &0.727 &0.774 \\
Diabetes &0.819 &0.6 \\
Cigarettes &0.766 &0.723 \\
Marijuana &0.589 &0.649 \\
Alcohol &0.683 &0.619 \\
\midrule
N &898 &3350 \\
\bottomrule
\end{tabular}
\end{table}


\begin{table}
    \centering
    \caption{Descriptive statistics for single-purchasers.}
    \label{tab:single_purchaser_demos_n}
    \footnotesize
    \begin{tabular}{p{1cm}llll}
\toprule
\multicolumn{2}{l}{\textbf{Attribute}} &\multicolumn{2}{l}{\textbf{n (N=898)}} \\\midrule
\multirow{2}{*}{\textbf{Gender}} & Male & 398 & \\
& Female & 471 & \\
\hline
\multirow{3}{*}{\textbf{Age}} &18 - 34 years &463 & \\
& 35 - 54 years &288 & \\
& 55 and older &147 & \\
\hline
\multirow{3}{*}{\textbf{Race}} & White &746 & \\
& Asian &77 & \\
& Black &74 & \\
\midrule
\multicolumn{5}{p{7.5cm}}{\emph{"Are any of the following the case for you or someone in your household or someone you share your Amazon account with?"}} \\
\hline
& &\textbf{Yes} &\textbf{No} \\
\hline
\textbf{Health}& Diabetes &61 &835 \\
\textbf{\&} & Cigarettes &98 &760 \\
\textbf{lifestyle}& Marijuana &160 &705 \\
& Alcohol &358 &496 \\
\bottomrule
\end{tabular}
\end{table}

There are two survey questions that enable analysis comparing "single" versus "non-single purchasers".

(1) How many people do you share your Amazon account with? i.e. how many people log in and make orders using your account?

(2) How many people are in your "household"?

The number of responses to these questions are shown in Figure~\ref{fig:amzn_account_sharing_counts}.
We call participants who answered "1 (just me!)" to these two questions "single purchasers". 
Single purchasers are a small subsample, n=898.

We then used the same test samples from the main analyses, but where data were further split based on whether users were single versus non-single purchasers.
We used the models that had already been trained on the full training data (training was not restricted to single versus non-single purchasers), to then predict the labels for the single and non-single test users. Results are shown in Table~\ref{tab:single_purchaser_auc_results}.  
Descriptive statistics for the single purchasers are shown in Table~\ref{tab:single_purchaser_demos_n}.

\subsection{Model performance increases with data}
\label{appendix:changes with data}

Tables \ref{tab:auc_results_by_cat_additions_rand_fraction_race}, \ref{tab:auc_results_by_cat_additions_rand_fraction_age}, \ref{tab:auc_results_by_cat_additions_rand_fraction_race}, and \ref{tab:auc_results_by_cat_additions_rand_fraction_health} show how model performance (AUC) increases with increasing fractions of purchase categories.
Table \ref{tab:auc_results_by_cat_acquisitions} shows how model performance changes with high the addition of categories chosen based on high profile Amazon market acquisitions.

\begin{table}
    \centering
    \caption{Increase in model performance (AUC) with increase in the fraction of randomly selected categories available to the model for training and testing: Gender.}
    \label{tab:auc_results_by_cat_additions_rand_fraction_gender}
    \footnotesize
    \begin{tabular}{lrrrrr}\toprule
\textbf{} &\multicolumn{2}{c}{\textbf{Male}} &\multicolumn{2}{c}{\textbf{Female}} \\\cmidrule{2-5}
\textbf{p } &\textbf{mean} &\textbf{95\% CI} &\textbf{mean} &\textbf{95\% CI} \\\midrule
\textbf{0.01} &0.61 &(0.59, 0.631) &0.62 &(0.588, 0.652) \\
\textbf{0.02} &0.67 &(0.643, 0.698) &0.684 &(0.656, 0.713) \\
\textbf{0.03} &0.709 &(0.688, 0.729) &0.731 &(0.709, 0.753) \\
\textbf{0.04} &0.728 &(0.712, 0.743) &0.755 &(0.738, 0.773) \\
\textbf{0.05} &0.741 &(0.726, 0.756) &0.768 &(0.755, 0.781) \\
\textbf{0.1} &0.792 &(0.777, 0.807) &0.823 &(0.81, 0.836) \\
\textbf{0.15} &0.819 &(0.809, 0.829) &0.848 &(0.841, 0.856) \\
\textbf{0.2} &0.839 &(0.829, 0.849) &0.863 &(0.855, 0.871) \\
\textbf{0.3} &0.86 &(0.856, 0.864) &0.88 &(0.874, 0.886) \\
\textbf{0.4} &0.871 &(0.865, 0.877) &0.892 &(0.888, 0.897) \\
\textbf{0.5} &0.878 &(0.874, 0.882) &0.899 &(0.896, 0.903) \\
\textbf{0.6} &0.883 &(0.879, 0.887) &0.902 &(0.899, 0.906) \\
\textbf{0.7} &0.888 &(0.884, 0.891) &0.906 &(0.904, 0.908) \\
\textbf{0.8} &0.891 &(0.888, 0.893) &0.909 &(0.906, 0.911) \\
\textbf{0.9} &0.891 &(0.89, 0.893) &0.911 &(0.909, 0.913) \\
\textbf{1} &0.893 & &0.912 & \\
\bottomrule
\end{tabular}
\end{table}

\begin{table*}[h]
    \centering
    \caption{Increase in model performance (AUC) with increase in the fraction of randomly selected categories available to the model for training and testing: Age.}
    \label{tab:auc_results_by_cat_additions_rand_fraction_age}
    \footnotesize
    \begin{tabular}{lrrrrrrr}\toprule
\textbf{} &\multicolumn{2}{c}{\textbf{18 - 34 years}} &\multicolumn{2}{c}{\textbf{35 - 54 years}} &\multicolumn{2}{c}{\textbf{55 and older}} \\\cmidrule{2-7}
\textbf{p } &\textbf{mean} &\textbf{95\% CI} &\textbf{mean} &\textbf{95\% CI} &\textbf{mean} &\textbf{95\% CI} \\\midrule
\textbf{0.01} &0.557 &(0.538, 0.576) &0.541 &(0.523, 0.56) &0.564 &(0.545, 0.583) \\
\textbf{0.02} &0.586 &(0.571, 0.6) &0.566 &(0.547, 0.585) &0.575 &(0.553, 0.597) \\
\textbf{0.03} &0.605 &(0.591, 0.619) &0.582 &(0.559, 0.606) &0.6 &(0.579, 0.621) \\
\textbf{0.04} &0.628 &(0.615, 0.641) &0.588 &(0.569, 0.607) &0.636 &(0.612, 0.661) \\
\textbf{0.05} &0.641 &(0.633, 0.649) &0.598 &(0.581, 0.614) &0.655 &(0.623, 0.687) \\
\textbf{0.1} &0.682 &(0.668, 0.696) &0.632 &(0.612, 0.652) &0.71 &(0.689, 0.731) \\
\textbf{0.15} &0.706 &(0.696, 0.716) &0.651 &(0.634, 0.668) &0.741 &(0.724, 0.758) \\
\textbf{0.2} &0.725 &(0.713, 0.736) &0.659 &(0.649, 0.669) &0.76 &(0.746, 0.773) \\
\textbf{0.3} &0.747 &(0.739, 0.756) &0.674 &(0.665, 0.683) &0.794 &(0.78, 0.808) \\
\textbf{0.4} &0.768 &(0.764, 0.773) &0.684 &(0.675, 0.692) &0.813 &(0.799, 0.827) \\
\textbf{0.5} &0.78 &(0.774, 0.787) &0.696 &(0.689, 0.702) &0.83 &(0.821, 0.838) \\
\textbf{0.6} &0.788 &(0.782, 0.794) &0.701 &(0.696, 0.707) &0.84 &(0.832, 0.848) \\
\textbf{0.7} &0.794 &(0.789, 0.8) &0.706 &(0.699, 0.712) &0.852 &(0.848, 0.856) \\
\textbf{0.8} &0.8 &(0.796, 0.804) &0.704 &(0.7, 0.708) &0.858 &(0.855, 0.861) \\
\textbf{0.9} &0.804 &(0.802, 0.806) &0.707 &(0.704, 0.71) &0.861 &(0.857, 0.866) \\
\textbf{1} &0.808 & &0.712 & &0.868 & \\
\bottomrule
\end{tabular}
\end{table*}

\begin{table*}
    \centering
    \caption{Increase in model performance (AUC) with increase in the fraction of randomly selected categories available to the model for training and testing: Race.}
    \label{tab:auc_results_by_cat_additions_rand_fraction_race}
    \footnotesize
    \begin{tabular}{lrrrrrrr}\toprule
\textbf{} &\multicolumn{2}{c}{\textbf{White}} &\multicolumn{2}{c}{\textbf{Asian}} &\multicolumn{2}{c}{\textbf{Black}} \\\cmidrule{2-7}
\textbf{p } &\textbf{mean} &\textbf{95\% CI} &\textbf{mean} &\textbf{95\% CI} &\textbf{mean} &\textbf{95\% CI} \\\midrule
\textbf{0.01} &0.564 &(0.539, 0.59) &0.539 &(0.515, 0.564) &0.562 &(0.539, 0.584) \\
\textbf{0.02} &0.59 &(0.565, 0.614) &0.536 &(0.516, 0.556) &0.58 &(0.561, 0.599) \\
\textbf{0.03} &0.613 &(0.597, 0.628) &0.566 &(0.541, 0.591) &0.605 &(0.588, 0.622) \\
\textbf{0.04} &0.629 &(0.609, 0.649) &0.579 &(0.558, 0.599) &0.621 &(0.599, 0.642) \\
\textbf{0.05} &0.64 &(0.62, 0.659) &0.595 &(0.573, 0.617) &0.628 &(0.607, 0.648) \\
\textbf{0.1} &0.681 &(0.668, 0.694) &0.645 &(0.63, 0.66) &0.673 &(0.656, 0.691) \\
\textbf{0.15} &0.706 &(0.697, 0.714) &0.673 &(0.657, 0.69) &0.707 &(0.694, 0.721) \\
\textbf{0.2} &0.719 &(0.712, 0.725) &0.68 &(0.662, 0.699) &0.716 &(0.703, 0.73) \\
\textbf{0.3} &0.736 &(0.729, 0.742) &0.71 &(0.694, 0.726) &0.741 &(0.729, 0.753) \\
\textbf{0.4} &0.751 &(0.739, 0.763) &0.726 &(0.71, 0.742) &0.76 &(0.751, 0.768) \\
\textbf{0.5} &0.761 &(0.75, 0.773) &0.737 &(0.726, 0.748) &0.76 &(0.75, 0.77) \\
\textbf{0.6} &0.766 &(0.756, 0.776) &0.742 &(0.731, 0.752) &0.774 &(0.762, 0.785) \\
\textbf{0.7} &0.77 &(0.761, 0.779) &0.747 &(0.738, 0.755) &0.784 &(0.773, 0.794) \\
\textbf{0.8} &0.78 &(0.772, 0.787) &0.754 &(0.748, 0.76) &0.788 &(0.779, 0.797) \\
\textbf{0.9} &0.788 &(0.786, 0.79) &0.762 &(0.76, 0.764) &0.802 &(0.796, 0.808) \\
\textbf{1} &0.791 & &0.763 & &0.806 & \\
\bottomrule
\end{tabular}
\end{table*}

\begin{table*}
    \centering
    \caption{Increase in model performance (AUC) with increase in the fraction of randomly selected categories available to the model for training and testing: Health/lifestyle.}
    \label{tab:auc_results_by_cat_additions_rand_fraction_health}
    \footnotesize
    \begin{tabular}{lrrrrrrrrr}\toprule
\textbf{} &\multicolumn{2}{c}{\textbf{Diabetes}} &\multicolumn{2}{c}{\textbf{Cigarettes}} &\multicolumn{2}{c}{\textbf{Marijuana}} &\multicolumn{2}{c}{\textbf{Alcohol}} \\\cmidrule{2-9}
\textbf{p } &\textbf{mean} &\textbf{95\% CI} &\textbf{mean} &\textbf{95\% CI} &\textbf{mean} &\textbf{95\% CI} &\textbf{mean} &\textbf{95\% CI} \\\midrule
\textbf{0.01} &0.517 &(0.503, 0.531) &0.549 &(0.502, 0.597) &0.537 &(0.51, 0.564) &0.519 &(0.499, 0.54) \\
\textbf{0.02} &0.51 &(0.495, 0.524) &0.6 &(0.566, 0.635) &0.547 &(0.526, 0.568) &0.528 &(0.503, 0.553) \\
\textbf{0.03} &0.511 &(0.493, 0.529) &0.615 &(0.589, 0.641) &0.556 &(0.538, 0.575) &0.545 &(0.53, 0.561) \\
\textbf{0.04} &0.52 &(0.498, 0.541) &0.621 &(0.598, 0.643) &0.56 &(0.541, 0.579) &0.557 &(0.541, 0.572) \\
\textbf{0.05} &0.536 &(0.518, 0.554) &0.628 &(0.611, 0.646) &0.556 &(0.537, 0.575) &0.566 &(0.553, 0.578) \\
\textbf{0.1} &0.543 &(0.534, 0.552) &0.651 &(0.638, 0.665) &0.577 &(0.556, 0.598) &0.583 &(0.573, 0.592) \\
\textbf{0.15} &0.551 &(0.535, 0.567) &0.666 &(0.649, 0.683) &0.582 &(0.565, 0.6) &0.593 &(0.583, 0.603) \\
\textbf{0.2} &0.549 &(0.531, 0.567) &0.679 &(0.664, 0.695) &0.597 &(0.575, 0.618) &0.595 &(0.581, 0.609) \\
\textbf{0.3} &0.559 &(0.542, 0.576) &0.692 &(0.681, 0.702) &0.603 &(0.587, 0.618) &0.608 &(0.595, 0.621) \\
\textbf{0.4} &0.565 &(0.554, 0.576) &0.708 &(0.699, 0.716) &0.613 &(0.599, 0.628) &0.614 &(0.605, 0.624) \\
\textbf{0.5} &0.574 &(0.563, 0.585) &0.709 &(0.699, 0.718) &0.618 &(0.606, 0.63) &0.616 &(0.607, 0.624) \\
\textbf{0.6} &0.578 &(0.567, 0.589) &0.712 &(0.702, 0.722) &0.622 &(0.615, 0.63) &0.621 &(0.614, 0.628) \\
\textbf{0.7} &0.593 &(0.58, 0.605) &0.719 &(0.712, 0.727) &0.624 &(0.615, 0.633) &0.625 &(0.62, 0.631) \\
\textbf{0.8} &0.599 &(0.59, 0.607) &0.724 &(0.717, 0.73) &0.632 &(0.625, 0.638) &0.626 &(0.622, 0.631) \\
\textbf{0.9} &0.605 &(0.601, 0.609) &0.728 &(0.723, 0.732) &0.631 &(0.628, 0.635) &0.628 &(0.624, 0.632) \\
\textbf{1} &0.612 & &0.728 & &0.634 & &0.631 & \\
\bottomrule
\end{tabular}
\end{table*}

\begin{table*}[]
    \centering
    \caption{Model results (AUC) with additional categories, starting with just books.}
    \label{tab:auc_results_by_cat_acquisitions}
    \footnotesize
\begin{tabular}{lrrrrr}\toprule
&\textbf{Books} &\textbf{With footwear} &\textbf{With Whole Foods} &\textbf{All products} \\\cmidrule{2-5}
\textbf{Total categories} &\textbf{190} &\textbf{207} &\textbf{316} &\textbf{1619} \\\cmidrule{1-5}
\textbf{Attribute} &\multicolumn{4}{c}{\textbf{AUC}} \\\midrule
\textbf{Male} &0.656 &0.766 &0.798 &0.893 \\
\textbf{Female} &0.634 &0.779 &0.817 &0.912 \\
\textbf{18 - 34 years} &0.66 &0.676 &0.703 &0.808 \\
\textbf{35 - 54 years} &0.627 &0.666 &0.656 &0.712 \\
\textbf{55 and older} &0.619 &0.587 &0.688 &0.868 \\
\textbf{White} &0.663 &0.652 &0.712 &0.791 \\
\textbf{Asian} &0.624 &0.72 &0.689 &0.764 \\
\textbf{Black} &0.579 &0.612 &0.702 &0.806 \\
\textbf{Diabetes} &0.527 &0.498 &0.55 &0.612 \\
\textbf{Cigarettes} &0.575 &0.596 &0.623 &0.728 \\
\textbf{Marijuana} &0.55 &0.575 &0.577 &0.634 \\
\textbf{Alcohol} &0.589 &0.579 &0.593 &0.631 \\
\bottomrule
\end{tabular}
\end{table*}

\clearpage

\section{Categories contributing to risk}
\label{appendix:categories contribution to risk}

\begin{figure}
    \centerline{\includegraphics[width=1.1\linewidth]{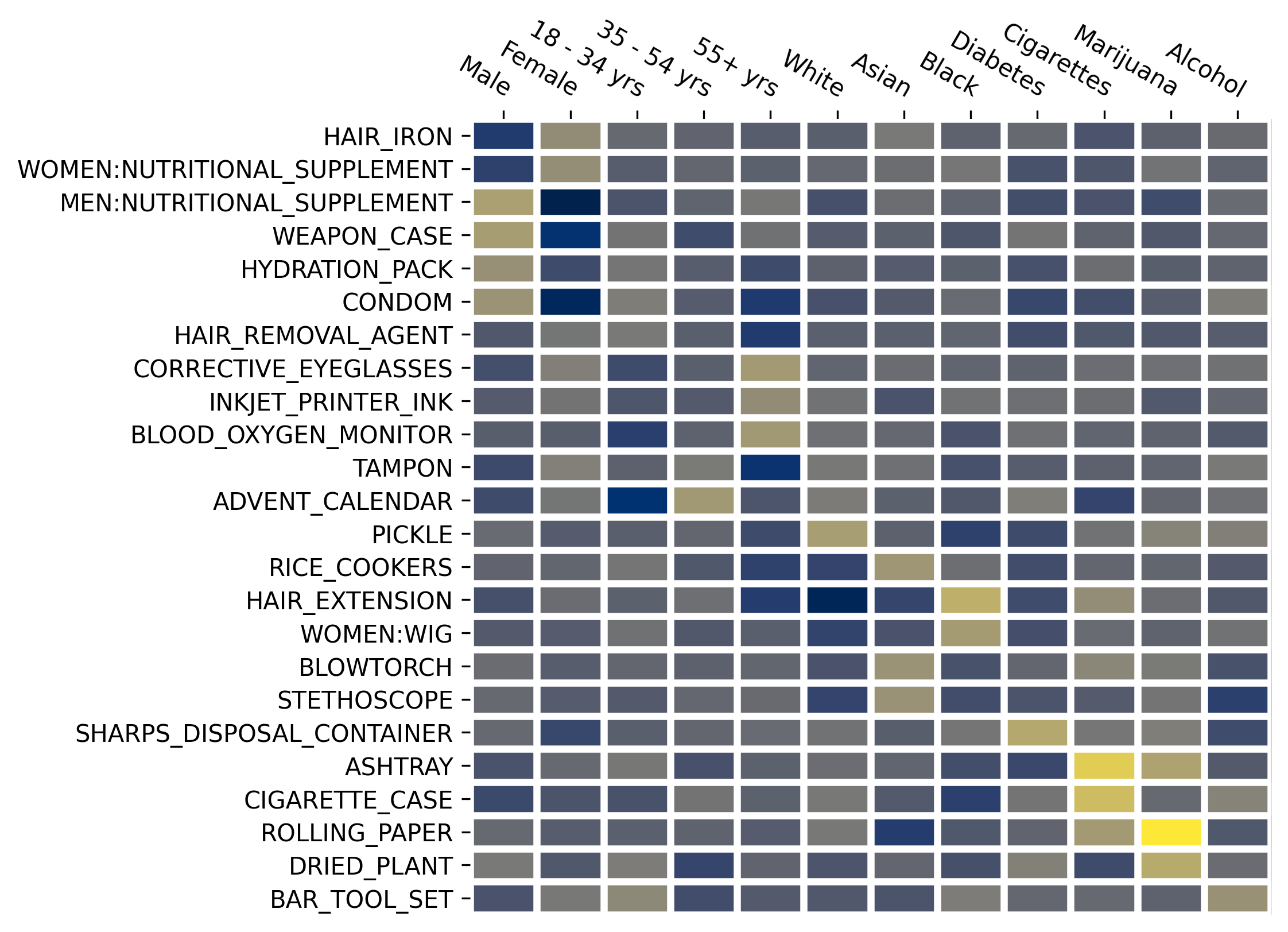}}
    \caption{Heatmap of ORs for categories contributing to inference risk, with all labels together.}
    \label{fig:categories_risk_all_together}
\end{figure}

\begin{table}
    \centering
    \caption{Results for the logistic regression versus SVM models, trained and tested on the same data as the main analysis.}
    \label{tab:lr_auc_results}
    \footnotesize
\begin{tabular}{lrrr}\toprule
\textbf{} &\multicolumn{2}{c}{\textbf{Model AUC}} \\\cmidrule{2-3}
\textbf{Attribute} &\textbf{Logistic regression} &\textbf{SVM} \\\midrule
\textbf{Gender} & & \\
Male &0.822 &0.893 \\
Female &0.852 &0.912 \\
\textbf{Age} & & \\
18 - 34 years &0.737 &0.808 \\
35 - 54 years &0.632 &0.712 \\
55 and older &0.802 &0.868 \\
\textbf{Race} & & \\
White &0.749 &0.791 \\
Black &0.769 &0.806 \\
Asian &0.74 &0.764 \\
\textbf{Health / lifestyle} & & \\
Diabetes &0.569 &0.612 \\
Cigarettes &0.681 &0.728 \\
Marijuana &0.618 &0.634 \\
Alcohol &0.618 &0.631 \\
\bottomrule
\end{tabular}
\end{table}

Tables \ref{tab:most_predictive_cats_gender}, \ref{tab:most_predictive_cats_age}, \ref{tab:most_predictive_cats_race}, \ref{tab:most_predictive_cats_health}
show the top 10 most positive and negative predictor categories for each attribute with ORs estimated via logistic regression.

Figure~\ref{fig:categories_risk_all_together} shows a heatmap of the ORs for the categories and attributes included in Figure~\ref{fig:categories_risk} all together.
The most predictive categories for cigarette and marijuana use stand out most brightly.

\begin{table*}
    \centering
    \caption{Most positive and negative predictor categories estimated via logistic regression: Gender.}
    \label{tab:most_predictive_cats_gender}
    \footnotesize
    \begin{tabular}{lrrrrr}\toprule
&\multicolumn{2}{r}{\textbf{Men}} &\multicolumn{2}{r}{\textbf{Female}} \\\cmidrule{2-5}
&\textbf{Categories} &\textbf{OR} &\textbf{Categories} &\textbf{OR} \\\midrule
\parbox[t]{2mm}{\multirow{10}{*}{\rotatebox[origin=c]{90}{Most positive predictors}}} &MEN:NUTRITIONAL\_SUPPLEMENT &6.09 &CELL\_PHONE\_HOLSTER &5.77 \\
&FAUCET\_WATER\_AERATOR &5.85 &LIGHT\_BOX &3.85 \\
&WEAPON\_CASE &5.52 &DISPLAY\_ENCLOSURE &3.83 \\
&WOMEN:NAIL\_POLISH &4.56 &DRAWER\_SLIDE &3.69 \\
&MEN:TOPICAL\_HAIR\_REGROWTH\_TREATMENT &4.46 &WOMEN:NUTRITIONAL\_SUPPLEMENT &3.65 \\
&CONDOM &4.27 &WASHER &3.57 \\
&HYDRATION\_PACK &3.89 &WOMEN:SHOE\_INSERT &3.55 \\
&SOLID\_FIRE\_FUEL &3.8 &HAIR\_IRON &3.48 \\
&MEN:VITAMIN &3.62 &SOAP\_DISH &3.41 \\
&BOTTLE\_RACK &3.49 &WOMEN:BOOK &3.38 \\
\midrule
\parbox[t]{2mm}{\multirow{10}{*}{\rotatebox[origin=c]{90}{Most negative predictors}}} &MEN:SLEEPWEAR &0.27 &CONTROLLER &0.26 \\
&VEHICLE\_ACCENT\_LIGHT &0.27 &PAW\_HOOF\_PROTECTOR &0.26 \\
&WOMEN:BOOK &0.27 &WRITING\_PAPER &0.25 \\
&CELL\_PHONE\_HOLSTER &0.26 &MEN:HAIR\_TRIMMER &0.23 \\
&WOMEN:APPAREL\_HEAD\_NECK\_COVERING &0.26 &MEN:SWIMWEAR &0.22 \\
&LIGHT\_BOX &0.26 &WEAPON\_CASE &0.22 \\
&FLAME\_WICK &0.25 &FAUCET\_WATER\_AERATOR &0.19 \\
&SOAP\_DISH &0.24 &WOMEN:NAIL\_POLISH &0.17 \\
&EXERCISE\_BLOCK &0.19 &CONDOM &0.16 \\
&WOMEN:SHOE\_INSERT &0.16 &MEN:NUTRITIONAL\_SUPPLEMENT &0.13 \\
\bottomrule
\end{tabular}
\end{table*}

\begin{table*}
    \centering
    \caption{Most positive and negative predictor categories estimated via logistic regression: Age.}
    \label{tab:most_predictive_cats_age}
    \footnotesize
    \begin{tabular}{lrrrrrrr}\toprule
&\multicolumn{2}{r}{\textbf{18 - 34 years}} &\multicolumn{2}{r}{\textbf{35 - 54 years}} &\multicolumn{2}{r}{\textbf{55 and older}} \\\cmidrule{2-7}
&\textbf{Categories} &\textbf{OR} &\textbf{Categories} &\textbf{OR} &\textbf{Categories} &\textbf{OR} \\\midrule
\parbox[t]{2mm}{\multirow{10}{*}{\rotatebox[origin=c]{90}{Most positive predictors}}} &MEN:KNIFE &4.59 &ADVENT\_CALENDAR &4.84 &CORRECTIVE\_EYEGLASSES &5.07 \\
&BRAKE\_ROTOR &4.47 &PROJECTILE\_BOW &4.71 &BLOOD\_OXYGEN\_MONITOR &4.83 \\
&BRAKE\_KIT &4.3 &AUTO\_BATTERY &4.06 &TOASTER &4.68 \\
&FISHING\_ROD\_REEL\_COMBO &4.12 &SNOW\_PANT &3.98 &WOMEN:APPAREL\_GLOVES &4.26 \\
&MEN:BASE\_LAYER\_APPAREL\_SET &4 &SHOWER\_CAP &3.74 &CLOTHES\_PIN &3.98 \\
&SOUS\_VIDE\_MACHINE &3.97 &INSTALLATION\_SERVICES &3.45 &MOP\_BUCKET\_SET &3.84 \\
&CHISEL &3.88 &POWER\_SUPPLIES\_OR\_PROTECTION &3.3 &BAKING\_PAN &3.7 \\
&WOMEN:HOME\_BED\_AND\_BATH &3.84 &MASSAGE\_STICK &3.26 &INKJET\_PRINTER\_INK &3.52 \\
&PALETTE\_PUTTY\_KNIFE &3.81 &MEN:WATCH\_BAND &3.25 &FISH &3.22 \\
&BOOK:CATERPILLARS &3.65 &ADULT\_COSTUME &3.16 &POPCORN\_POPPER &3.12 \\
\midrule
\parbox[t]{2mm}{\multirow{10}{*}{\rotatebox[origin=c]{90}{Most negative predictors}}} &CRIB &0.26 &BOTTLE\_RACK &0.31 &COSMETIC\_SPONGE &0.3 \\
&BOOK:TRAVEL &0.26 &BOOK:LAW &0.31 &HAIR\_REMOVAL\_AGENT &0.29 \\
&WOMEN:CORRECTIVE\_EYEGLASSES &0.24 &CARRYING\_CASE\_OR\_BAG &0.3 &PARTY\_DECORATION\_PACK &0.29 \\
&ZIPPER\_FASTENER &0.23 &WATER\_DISPENSER &0.3 &VEHICLE\_LIGHT\_ASSEMBLY &0.29 \\
&WATER\_HEATER &0.23 &BLEACH &0.29 &CONDOM &0.28 \\
&POWER\_SUPPLIES\_OR\_PROTECTION &0.22 &MEN:BASE\_LAYER\_APPAREL\_SET &0.29 &VIDEO\_GAME\_CONTROLLER &0.27 \\
&MEN:APPAREL\_GLOVES &0.22 &BOOK:CATERPILLARS &0.29 &DOWNLOADABLE\_VIDEO\_GAME &0.24 \\
&ADVENT\_CALENDAR &0.21 &VEHICLE\_EMBLEM &0.28 &PIERCING\_JEWELRY &0.24 \\
&SURVIVAL\_KIT &0.21 &CHAINSAW\_CHAIN &0.16 &TAMPON &0.23 \\
&AV\_RECEIVER &0.2 &LIQUID\_FUEL\_CONTAINER &0.14 &GARMENT\_STEAMER &0.22 \\
\bottomrule
\end{tabular}
\end{table*}

\begin{table*}
    \centering
    \caption{Most positive and negative predictor categories estimated via logistic regression: Race.}
    \label{tab:most_predictive_cats_race}
    \footnotesize
    \begin{tabular}{lrrrrrrr}\toprule
&\multicolumn{2}{c}{\textbf{White}} &\multicolumn{2}{c}{\textbf{Asian}} &\multicolumn{2}{c}{\textbf{Black}} \\\cmidrule{2-7}
&\textbf{Categories} &\textbf{OR} &\textbf{Categories} &\textbf{OR} &\textbf{Categories} &\textbf{OR} \\\midrule
\parbox[t]{2mm}{\multirow{10}{*}{\rotatebox[origin=c]{90}{Most positive predictors}}} &WOMEN:KEYCHAIN &7.51 &MUSICAL\_INSTRUMENT\_TOY &6.62 &HAIR\_EXTENSION &9.07 \\
&PICKLE &5.61 &PUNCHING\_BAG &6.09 &EXERCISE\_STEP\_PLATFORM &7.74 \\
&FACE\_SHAPING\_MAKEUP &5.1 &ABDOMINAL\_EXERCISER &5.74 &WOMEN:WAIST\_CINCHER &6.29 \\
&BOOK:ANTIQUES \& COLLECTIBLES &5.09 &WOMEN:BREAST\_PETAL &5.17 &WOMEN:WIG &5.37 \\
&CARBURETOR &4.81 &RICE\_COOKERS &4.58 &HAIR\_STYLING\_AGENT &5.28 \\
&TRAMPOLINE &4.44 &BLOWTORCH &4.24 &INSTALLATION\_SERVICES &4.38 \\
&SWIMWEAR &4.25 &BISS &4.24 &WOMEN:BODY\_POSITIONER &4.11 \\
&WOMEN:NECKLACE &4.11 &STETHOSCOPE &4.05 &WOMEN:SHOWER\_CAP &3.56 \\
&BOOK:NATURE &4.02 &SHOE\_TREE &4.02 &ELECTRONIC\_SENSOR &3.56 \\
&FIREPLACE &4.01 &ARM\_SLEEVE &3.67 &ROASTING\_PAN &3.48 \\
\midrule
\parbox[t]{2mm}{\multirow{10}{*}{\rotatebox[origin=c]{90}{Most negative predictors}}} &WOMEN:SWEATBAND &0.27 &ROLLING\_PAPER &0.3 &JERKY &0.32 \\
&WATERING\_CAN &0.27 &PHYSICAL\_TV\_SERIES &0.3 &CELL\_PHONE\_HOLSTER &0.3 \\
&NON\_DAIRY\_CREAM &0.26 &MEN:CONDITIONER &0.3 &TOASTER &0.3 \\
&ELECTRONIC\_COMPONENT\_TERMINAL &0.25 &ANIMAL\_CAGE &0.3 &MEN:SWIMWEAR &0.29 \\
&COUNTERTOP\_OVEN &0.24 &COUNTERTOP\_BURNER &0.29 &VIDEO\_CARD &0.27 \\
&MUSICAL\_INSTRUMENT\_TOY &0.23 &TOASTER &0.28 &PERSONAL\_CARE\_APPLIANCE &0.27 \\
&TONG\_UTENSIL &0.2 &POPCORN\_POPPER &0.28 &WIPER\_BLADE &0.23 \\
&HAIR\_EXTENSION &0.15 &MEN:HAIR\_COMB &0.27 &SEWING\_BUTTON &0.22 \\
&SOLAR\_PANEL &0.14 &MOUNT\_BRACKET &0.25 &WILDLIFE\_FEEDER &0.21 \\
&WOMEN:WAIST\_CINCHER &0.13 &SAFE &0.25 &DISHWASHER\_DETERGENT &0.2 \\
\bottomrule
\end{tabular}
\end{table*}

\begin{table*}
    \centering
    \caption{Most positive and negative predictor categories estimated via logistic regression: Health and lifestyle.}
    \label{tab:most_predictive_cats_health}
    \footnotesize
    \begin{tabular}{lrrrr}\toprule
&\multicolumn{2}{r}{\textbf{Diabetes}} &\multicolumn{2}{r}{\textbf{Cigarettes}} \\\cmidrule{1-5}
&\textbf{Categories} &\textbf{OR} &\textbf{Categories} &\textbf{OR} \\\midrule
\parbox[t]{2mm}{\multirow{10}{*}{\rotatebox[origin=c]{90}{Most positive}}} &SHOWER\_CAP &8.52 &ASHTRAY &18.21 \\
&SHARPS\_DISPOSAL\_CONTAINER &7.45 &CIGARETTE\_CASE &12.71 \\
&ANIMAL\_MUZZLE &6.75 &UNCATEGORIZED &6.2 \\
&GARDEN\_TOOL\_SET &6.01 &MEN:SPORT\_ACTIVITY\_GLOVE &5.89 \\
&DISPOSABLE\_INCONTINENCE\_SURFACE\_PROTECTOR &5.59 &VEHICLE\_COVER &5.31 \\
&INDUSTRIAL\_CASTERS &5.2 &ROLLING\_PAPER &5.07 \\
&MAGAZINES &4.78 &MOTHERBOARD &5.01 \\
&CLOTHES\_PIN &4.59 &BOOK:DRAMA &4.99 \\
&SPOON\_REST &4.51 &MEN:BACKPACK &4.98 \\
&UTILITY\_MAGNET &4.49 &SHEET\_PAPER &4.75 \\
\midrule
\parbox[t]{2mm}{\multirow{10}{*}{\rotatebox[origin=c]{90}{Most negative}}} &COLLAPSIBLE\_PLAY\_STRUCTURE &0.24 &BOOKMARK &0.24 \\
&WATER\_DISPENSER &0.23 &DAIRY\_BASED\_ICE\_CREAM &0.24 \\
&FOOD\_SHAKER &0.23 &BLUE\_LIGHT\_BLOCKING\_EYEGLASSES &0.23\\
&WOMEN:HAIR\_TRIMMER &0.22 &TOOTHBRUSH\_HEAD &0.23 \\
&LEAVENING\_AGENT &0.2 &WOMEN:SKIN\_MOISTURIZER &0.23 \\
&BAKING\_STONE &0.19 &WIG &0.22 \\
&DOOR\_CLOSER &0.17 &WOMEN:SNOW\_PANT &0.21 \\
&ELECTRONIC\_COMPONENT &0.15 &SPINNING\_TOY\_TOP &0.19 \\
&BREAST\_PETAL &0.14 &TOY\_MODEL\_VEHICLE\_TRACK &0.18 \\
&EXERCISE\_BLOCK &0.14 &BODY\_CARE\_PRODUCT &0.17 \\
\midrule
&\multicolumn{2}{r}{\textbf{Marijuana}} &\multicolumn{2}{r}{\textbf{Alcohol}} \\\cmidrule{1-5}
&\textbf{Categories} &\textbf{OR} &\textbf{Categories} &\textbf{OR} \\\midrule
\parbox[t]{2mm}{\multirow{10}{*}{\rotatebox[origin=c]{90}{Most positive}}} &ROLLING\_PAPER &34.15 &SNOW\_PANT &4.47 \\
&DRIED\_PLANT &7.85 &VEHICLE\_WRAP &4.1 \\
&ASHTRAY &6.52 &BAR\_TOOL\_SET &3.97 \\
&TOGGLE\_SWITCH &5.9 &HIP\_FLASK &3.81 \\
&VEHICLE\_BUMPER &4.2 &WOMEN:CANDLE &3.78 \\
&ALARM &4.16 &DISHWASHER &3.53 \\
&FACIAL\_TISSUE\_HOLDER &3.85 &EGG &3.35 \\
&EXERCISE\_STRAP &3.64 &COOKING\_CONTAINER &3.32 \\
&IGNITION\_COIL &3.62 &BOTTLE\_RACK &3.27 \\
&FASHIONNECKLACEBRACELETANKLET &3.59 &MEN:DRINKING\_CUP &3.21 \\
\midrule
\parbox[t]{2mm}{\multirow{10}{*}{\rotatebox[origin=c]{90}{Most negative}}} &SHEET\_PAN &0.26 &MEN:HANDBAG &0.33 \\
&WOMEN:PROTECTIVE\_GLOVE &0.26 &VEHICLE\_WIND\_DEFLECTOR &0.32 \\
&WOMEN:LEG\_SLEEVE &0.26 &BUCKLE &0.32 \\
 &BREAD\_MAKING\_MACHINE &0.26 &BODY\_CARE\_PRODUCT &0.32 \\
&DECORATIVE\_POM\_POM &0.26 &DRIP\_IRRIGATION\_KIT &0.3 \\
&RECREATION\_BALL &0.25 &NAVIGATION\_COMPASS &0.29 \\
&TARP &0.24 &AV\_FURNITURE &0.25 \\
&SHOE\_ACCESSORY &0.24 &SPINNING\_TOY\_TOP &0.25 \\
&PILLOW\_PROTECTOR &0.21 &MESS\_KIT &0.24 \\
&LIGHT\_SOURCE &0.17 &BOOK:ENGLISH LANGUAGE &0.14 \\
\bottomrule
\end{tabular}
\end{table*}

\end{document}